\documentclass{ws-jai}
\pdfoutput=1
\usepackage{xspace}
\usepackage{aas_macros}
\begin{document}
\catchline{}{}{}{}{} 
\markboth{Buscher et al.}{Conceptual design of the MROI}

\title{THE CONCEPTUAL DESIGN OF THE MAGDALENA RIDGE OBSERVATORY
INTERFEROMETER}

\author{David F. Buscher$^1$, Michelle Creech-Eakman$^2$, Allen Farris$^2$,
Christopher A. Haniff$^{3}$, and John S. Young$^{3}$}
\address{
$^1$Cavendish Laboratory, University of Cambridge, J J Thompson Avenue, Cambridge, UK, 
dfb@mrao.cam.ac.uk\\ 
$^2$MRO, New Mexico Institute of Mining and Technology, 801 Leroy Place, Socorro, NM 87801, USA,
$^3$Cavendish Laboratory, University of Cambridge, J J Thompson Avenue, Cambridge, UK.
}

\maketitle

\footnotetext[1]{Corresponding author.}

\begin{history}
\received{(to be inserted by publisher)};
\revised{(to be inserted by publisher)};
\accepted{(to be inserted by publisher)};
\end{history}
\newcommand{\uv}{\ensuremath{(u,v)}\xspace}
\newcommand{\micron}{\,\ensuremath{\mu}{\rm m}\xspace}
\newcommand{\etc} {etc.\xspace}
\newcommand{\eg} {e.\,g.\xspace}
\newcommand{\ie} {i.\,e.\xspace}
\newcommand{\ii}{{\rm i}} 
\newcommand{\degree}{\nobreak\ensuremath{^\circ}\xspace}    
\newcommand{\etal}{{\it et~al.\/}}    
\newcommand{\simlt}{\stackrel{<}{_{\sim}}}
\newcommand{\simgt}{\stackrel{>}{_{\sim}}}

\begin{abstract}
We describe the scientific motivation for and conceptual design of the
Magdalena Ridge Observatory Interferometer, an imaging interferometer
designed to operate at visible and near-infrared wavelengths. The rationale for the major technical decisions in the interferometer design is discussed, the success of the concept is appraised, and the implications of this analysis for the design of future arrays are drawn out.
\end{abstract}


\section{Introduction} 
\label{sec:introduction}
The last 50 years have seen a drive to enhance the
angular resolution delivered by ground-based telescopes in the
presence of atmospheric turbulence. Most of this effort has focused on
the use of adaptive optics (AO): modern implementations can routinely
deliver 100 milliarcsecond diameter point spread functions. Nevertheless, the desire for angular resolutions better than this remains a
powerful driving force in astrophysics. For example, over half of the
key science cases comprising the E-ELT design reference mission \cite{liske11} require the diffraction-limited performance at the 10\,mas
level to be delivered.

Ground-based interferometric arrays have contributed
to studies on {\em much smaller} angular scales. Typical recent
studies have focused on Cepheid multiplicity \citep{gallenne_multiplicity_2013,li_causi_binarity_2013}, circumstellar disk physics \citep{kreplin_revealing_2013}, the orbital
dynamics of close binaries \citep{hummel_dynamical_2013,sana_three-dimensional_2013} and the fundamental parameters of single stars \citep{klotz_catching_2013,arroyo-torres_atmospheric_2013}. Interferometric arrays currently offer the only direct means to
study astrophysical phenomena on such small angular scales. However, the scientific output of the current
generation of ground-based arrays appears to have converged
to about 50 peer reviewed papers per year, focused almost exclusively
on galactic astrophysics.  This level and focus of activity has been
maintained for at least five years now, and this might suggest that
these major facility arrays are being constrained by the capabilities
of their initial suite of instruments.

The principal shortcomings of current interferometers are easily
identifiable and are broadly speaking poor sensitivity and the
inability, in most implementations, to routinely image targets with
high fidelity. These issues formed the
basis of the initial planning of the technical architecture of the
Magdalena Ridge Observatory Interferometer (MROI). The MROI was
conceived as part of an ambitious plan at the New Mexico Institute of
Mining and Technology (NMT) to develop a new astronomical observatory
at a high altitude site in central New Mexico to support a broad
portfolio of programmes in research, education, and the commercial and
defence arenas.

In this paper we review the conceptual design of the MROI, currently under 
construction on the Magdalena Ridge at an elevation of 10,500\,ft. From conception, the MROI was targeted towards 
expanding the capabilities of ground-based optical/near-infrared 
interferometry in three main directions: sensitivity, imaging capability, 
and speed of operation. However, it aimed to do this not by solving a set of extreme technical challenges, but 
rather by leveraging the many technical lessons that had been learnt 
developing the first generation of prototype interferometers and by 
capitalising on advances in technology that had taken place since 
those first arrays had been commissioned. In this sense it has always been 
a low-risk endeavour.  More importantly, though, its architecture was 
guided through an iterative process whereby the scientific ambitions of 
its proposed user community were moderated by transparent presentations of 
the technical constraints implicit in interferometric imaging and the 
current state-of-the-art in the management of the turbulent 
atmosphere.

The layout of our manuscript is as follows. We begin with a review of the 
limitations of existing arrays in section~\ref{sec:limitations} and follow 
this with a brief resum{\'e} of the broad areas of science the MROI has 
been designed to tackle in section~\ref{sec:science-drivers}. We then walk 
through the major technical requirements in section~\ref{sec:techn-requ} and the key design choices and trades that were made as the overall 
architecture of the MROI was frozen (section~\ref{sec:conceptual-design}). We present the high-level error budget for the array in 
section~\ref{sec:error-budget} and evidence that we can meet the key sensitivity requirement in section~\ref{sec:limiting-magnitude}. The main technical lessons learned during 
the design, prototyping and on-going deployment of the MROI are summarised 
in section~\ref{sec:lessons-learned} and our conclusions appear in 
section~\ref{sec:conclusions}.

\section{Limitations of existing interferometers}
\label{sec:limitations}

The scientific capabilities of conventional optical/IR telescopes are 
frequently characterised by their ``performance'' measured against half a 
dozen or so key metrics. These usually include such items as spatial resolution, spectral resolving power, wavelength range, sensitivity, field-of-view and so on. Almost all of these are appropriate figures of merit for interferometric 
arrays, but as written they ``hide'' perhaps the most important 
shortcoming of contemporary optical/IR arrays, that is, the fact that few 
are able to provide images of the targets under study. In the following 
sub-sections we discuss the background to this problem as well as the 
capabilities of existing interferometric arrays with reference to two of 
the more conventional performance metrics listed above: sensitivity and 
spatial resolution.

\subsection{Imaging capability}
\label{subsec:imaging}

The ability of an interferometric array to recover an image is limited 
fundamentally by how many independent measurements of the visibility 
function of the source, i.e.~the Fourier transform of the sky brightness 
distribution, can be secured. If a total of $N_{uv}$ independent 
visibility data are collected, this implies that an image with of order 
$N_{uv}$ degrees-of-freedom can be reconstructed. Thus, for example, if a 
$5 \times 5$ pixel image of an extended source is required, of order 25 
independent visibility data need to be collected. In addition, the 
projected baselines used to secure these data should range in length by a 
factor of roughly 5:1.

Meeting these requirements is a significant challenge for all existing 
optical/IR arrays. For example, at the CHARA array on Mt Wilson, the 
typical range in baseline lengths realised in imaging observations to date 
has been between 3 and 4:1 \citep[see e.g.][]{baron_imaging_2012} At the VLTI, baseline length ratios of up to 6:1 have been 
achieved but only by combining measurements made over several nights 
\cite{millour_imaging_2011} during which time some of the interferometer 
elements have been relocated. An additional issue is the long time needed 
to secure individual visibility ``snapshots''. At the VLTI, the 
allocations of time recommended for securing calibrated snapshot 
visibility data with the AMBER (3 visibilities) and MIDI (2 visibilities) 
instruments are between 50 and 90 minutes.

As a result of these issues, in excess of 90\% of the published scientific 
results from optical/IR interferometers rely upon fitting either geometric 
or physical models to the measured visibility data. In cases where the 
systems under study are well known, in the sense that they can be {\em 
reliably} characterised by a model with a small number of 
degrees-of-freedom, interferometric data can provide very powerful diagnostics 
for the precise values of the model parameters. However, there are many cases where the models 
themselves are in question, e.g. ``Is the dust distributed in a continuous 
disk or might an Archimedean spiral be preferred?'', or where the physical 
model used contains many parameters whose interaction is 
complex --- an example of such a physical model might be that of a 
dusty disk with an inner and outer radius, an inclination, parameters 
describing the radial and vertical scale-lengths, a dust mass, a dust 
composition, a grain size distribution and a stellar temperature and 
luminosity. In these cases the inability to recover an image assuming no preferred 
structure (\ie ``model-independent imaging'') is a major shortcoming of contemporary interferometers.

\subsection{Limiting sensitivity}
\label{subsec:sensitivity}

An almost equally pressing issue for the current generation of optical/IR 
interferometric arrays is their relatively poor sensitivity. Because 
ground-based arrays are constrained to operate in the presence of 
atmospheric fluctuations, they are fundamentally limited by the small 
values of Fried's parameter, $r_0$, and the atmospheric coherence time. 
$t_0$. In view of this, one might expect the limiting sensitivity of an 
optical/IR array to be similar to that of a natural guide star (NGS) AO 
system, i.e.~to be based on the brightness of the star being used to sense 
the atmospheric wavefront perturbations. At typical NGS AO systems 
reference stars as faint as $m_V \sim 12$ can used very effectively, and 
partial AO correction can still be realised with guide stars as faint as 
$m_V \sim 16$.

In practice, all existing optical/IR arrays struggle to come even close to 
these sensitivity limits. The faintest optical measurements secured at the 
NPOI, SUSI and CHARA arrays have been from targets with visual magnitudes 
in the range 5--7, while in the near-infrared $K$~band typical limiting 
magnitudes have been in the range 6--8. In a few cases fringe data have 
been secured on targets as faint as $m_K \sim 10$ but these studies have 
relied upon arrays that have exploited AO-corrected 8\,m-class telescopes 
as the basis array elements.

It is this relatively poor sensitivity that means that contemporary 
optical/IR interferometry has has such a small impact on extra-galactic 
astrophysics, and that most interferometric studies have been directed 
towards a small number of the brightest exemplars of different classes of 
targets. The ability to reach a limiting sensitivity comparable to that of 
current NGS AO installations would be a major advance.

\subsection{Angular resolution}
\label{subsec:angular-resolution}

A dimension along which existing interferometric implementations have made 
significant inroads is that of angular resolution. As mentioned in 
section~\ref{sec:introduction} above, all existing separated element 
optical/IR arrays have baselines sufficiently long --- typically at least 
100\,m --- that sensitivity to sub-10\,mas angular scales is guaranteed. 
There seems little doubt that arrays with baselines at least five times 
longer than this, i.e.~of order 500\,m are desirable. Baldwin and Haniff 
(2002) provide a useful table summarising the characteristic angular 
scales expected for a range of targets at distances which would ensure 
that more than only the brightest targets would be observable. Many of 
these classes of object, e.g.~main sequence stars, stellar gas shells, 
spectroscopic and interacting binaries, and AGN broad line regions, would 
only be expected to be resolved on baselines in excess of 300\,m.

What is less clear is whether arrays with baselines in excess of a 
kilometer are of high priority. Targets requiring such long baselines to 
resolve them must necessarily be small, and if so, they must be distant. 
In that case, it is likely they would be faint unless they had 
particularly high brightness temperatures. \citet{baldwin_application_2002} 
discuss this argument further, but notwithstanding the details, it seems 
probable that enhancing the imaging capability and sensitivity of existing 
arrays ought to take priority over extending their maximum baselines.

\section{Science drivers for the MROI}
\label{sec:science-drivers}

The science case for the MROI is predicated on three key areas,
familiar to many existing interferometric facilities today.  This is
really a reflection of the brightness temperatures of the targets and
the operational wavelengths and baselines accessible at most modern
facilities.  The key differences associated with the interferometric
observations at MROI, however, are in more accurate imaging and in
much greater sensitivities, both of which will dramatically change the
fundamental scientific questions that can be answered.  Our three key
areas, outlined briefly below, encapsulate the scientific drivers
which drive the design choices for the facility.

\subsection{AGN Astrophysics}
Though active galactic nuclei (AGN) have been studied for many years,
interferometric imaging offers the prospect of gaining important new
insights into their structure. The physical scales of most interest
include the broad line region (BLR) on sub 0.1 pc scales; the narrow
line region (NLR) extending from 1--1000\,pc and the dust torus which
ranges in size from $\sim 0.5-10$\,pc. In nearby AGN ($z<0.01$) these
correspond to angular sizes from 0.5\,mas to several seconds of arc,
but it is on scales smaller than 0.1 arcseconds where observations
with the MROI will have most leverage.

The limiting sensitivity of modern interferometers today means that
only about 1 dozen AGN have been studied at near or mid-infrared
wavelengths \citep[c.f. ][]{tristam09,honig12,kishimoto11}.  However, even this small quantity of data has been
intriguing. For instance, in the mid-infrared, Centaurus A appears to
have a thin, dusty disk, the axis of which may align with its radio
jet \citep{meisenheimer07} while NGC 3783 and the Circinus galaxy
appear clumpy \citep{beckert08,tristam07} and NGC 1068
may have a torus and funnel structure \citep{raban09}. Studying a
significantly larger sample of these galaxies with an interferometer
with complex imaging capabilities is required to understand the origin
and ubiquity of these morphologies.

A key goal for MROI will be to investigate the details and reliability
of dust torus models. It is being designed to address questions such
as: a) the frequency of occurrence of tori, b) the geometric and
physical properties of the obscuring material, and c) whether these
are consistent with ``unified'' schemes.  In particular,
model-independent images are crucial in interpreting observations of
clumpy (but small) tori, which are not well-described by a small
number of parameters but which are increasingly favoured. Simulations
of the archetype NGC 1068 \citep{honig_radiative_2006} predict K-band
visibilities greater than 30\% on baselines of up to 50\,m, a regime
where the most compact configurations of the MROI (with baselines from
8--45\,m) will be ideally matched to the relevant scales. Furthermore it
should be possible to correlate the torus axis with the larger scale
radio emission.

A more ambitious study to be undertaken with MROI will be to
investigate the BLR/NLR transition region. Imaging the transition
region between the BLR and the NLR offers the prospect of isolating
the outflowing line-emitting gas and in turn determining the origin,
geometric, physical and temporal characteristics of the outflow. This
would result in a breakthrough in the understanding of the dynamics of
AGN cores, of which little is currently known.

\subsection{YSO Astrophysics}
A second major theme for MROI will be to advance detailed studies of
star and planet formation through studies of young stellar objects
(YSOs). As with the AGN, results from first-generation near-IR
interferometers have led to significant new insights into the complex
inner structure of protostellar/planetary disks. For example, the
sizes of Herbig Ae/Be and T Tauri objects have been measured to be 3-7
times larger than predicted by geometrically-thin disk models
previously used to explain the SED measurements \citep[see e.g.][and references therein]{millan07}, and have led to the
development of a new class of ``puffed up'' models for the flared disk
emission \citep[e.g.][]{dullemond01,isella05}.  Other
measurements have produced tantalising evidence for structures in the
inner disk relevant to the planet formation process. For example,
hotspots have been detected in the disks of AB Aur \citep{millan-gabet_bright_2006}  and FU Orionis \citep{malbet05}, and evidence for gap
clearing has been found in LkCa 15 \citep{espaillat08}.

What has been missing in all of these studies has been sufficient
visibility and closure phase data to properly constrain the complex
geometrical and physical structure of the inner, \ie sub-10-AU-scale,
regions of the disk. Measurements of order one hundred Fourier
amplitudes and closure phases, at sensitivity levels a several
magnitudes fainter than what is currently possible, will be critical
to understanding the physical processes taking place there.

A key goal for MROI's early science will be to provide a census of
disk properties for well-defined samples of low, intermediate and
high-mass stars. By allowing hundreds of targets to be observed with
excellent Fourier plane coverage, it will be possible to critically
assess theoretical models which predict not only the structure
expected but also the temporal evolution of the dust. This evolution
is expected to be strongly impacted by the presence of any planetary
or brown-dwarf companions in the inner disk, which should lead to
either disk breakup or disk-clearing \citep{creech10}; such
structures will certainly be detectable in active YSO disks with
MROI. The higher resolution spectral modes of MROI ($R \sim 300$) will allow
it to discriminate between dust, gas and molecular emission using
spectral diagnostics such as the Brackett gamma line and CO bandheads
\citep[c.f.][]{eisner07}. MROI will thus enable routine monitoring of
the dust and gas on scales that have Keplerian rotational timescale of
weeks to years.

A parallel theme will be to search for low-mass companions in star
forming regions, not only from the perspective of precise dynamical
mass and age estimation \citep[e.g.][]{konopacky07a} but also to
validate models of star formation. To date this has been undertaken
using high resolution imaging methods on single telescopes
\citep[e.g. shift-and-add imaging, speckle interferometry or AO --- see][]{konopacky07b}, but these are limited in both angular
resolution and the ability to detect the lowest-mass companions. MROI,
when employed in the compact configuration, will be ideally suited to
make major inroads in this area: companions with a K-band flux 1/100
that of the primary star will be detectable within a
night's observations, allowing numerous candidates in, e.g., Taurus
and Orion to be surveyed to well below the hydrogen burning limit.

\subsection{Mass-Loss and Dynamical Systems}

The third key area for optical/infrared interferometry is the study of
fundamental physical processes.  Unique contributions can be made by
imaging interferometers with sub-milliarcsecond resolution whenever
complex systems/processes are being studied which are difficult to
understand without direct imaging.  In this case, MROI will focus on
the physics of mass loss, mass transfer in binaries, and
time-domain/dynamical interactions.  While this type of work is in
principle possible for any interferometer, it becomes much more
powerful with an increase in the number of apertures (and consequent numbers of
visibilities and closure-phases) that are used and with shorter
timescales for producing complete images.  Excellent recent examples
of images in these broad scientific areas include: CHARA imaging
results on the long-period eclipsing system, Epsilon Aurigae \citep{kloppenborg_infrared_2010}, and on the interacting binary system Beta
Lyrae \citep{zhao08} and older aperture masking results from the
Keck Observatory on interacting binary/mass-losing system WR104
\citep{tuthill08}.

Because few actual images exist today, except for the brightest
``archetypes'' of each class of object, it is expected that tremendous
insight into all the evolutionary stages of low to high-mass stars
will be accomplished when statistical samples of objects in each class
are accumulated.  Some of the questions MROI will address include: a) how does the pulsational behaviour in the late
stages of life of an AGB star affect the mass-loss and subsequent
shaping processes of planetary nebulae; b) what are the
characteristics of mass-loss processes at different stages in stellar
evolution, i.e. continuous, episodic, clumpy, smooth; c) what is the
interaction between the stellar surface (\ie magnetic fields, star
spots) and the mass-loss processes as traced over a variety of
time-scales; d) how does mass-transfer in interacting systems trigger
subsequent explosive events; e) what is the connection between stellar
``shape'' (\ie non-spherical, rapidly rotating systems) and their wind
structures; and f) do optical or infrared counterparts exist to trace
known phenomena at X-ray/UV wavelengths, for example in the shocks or
hot winds, for high-mass systems?  Papers which shed some light on
these questions have been published in the last few years and include:
on the nova RS Ophiuchi \cite{barry_milliarcsecond_2008}, on star-spots on
Betelgeuse \citep{chiavassa10}, and on imaging of rapidly rotating
stars \citep[c.f.][and references therein]{vanbelle12}.  In all cases,
high-resolution images on statistical numbers of objects, and
especially over a variety of timescales, will only increase our
understanding of the physical processes involved.  MROI's design is
uniquely developed to address many of these questions.

\section{Technical requirements}
\label{sec:techn-requ}
In order to clarify the goals for the conceptual design of the MROI,
the functional and performance requirements for the array need to be
defined. The major requirements are briefly summarised below, but this
is only part of a much larger list. The requirements generally result
from a desire to address the three primary science missions outlined
above, i.e.~studies of active galactic nuclei, star and planet
formation, and stellar accretion and mass loss, but is important to
recognise that the requirements do not solely flow top-down from the
science case but result from a balancing scientific desirability and
technical feasibility: it will be seen below that many scientifically
desirable specifications conflict with one another when it comes to
implementing them together in a single design.

\subsection{Imaging}
\label{sec:imaging}
The attainment of a unique imaging capability was one of the chief
scientific drivers for the MROI. The imaging performance can be
characterised crudely by the number of resolution elements (often
called ``resels'' in analogy to pixels) in an image but for
interferometric imaging the number of resels which can be derived is
dependent not only on the angular size of the object but also on the
object morphology. Objects such as stellar disks have visibility
functions which fall rapidly from unity with increasing spatial
frequency. On the baselines which have sufficient angular resolution
to see interesting features such as surface activity, the majority of
the flux in the object is resolved and the visibility can be so low
that the signal-to-noise ratio of the fringes is below the level
required to find the interferometric fringes even on bright
targets. Our experience with the COAST interferometer was that many
such objects could not be observed with more than a few resels across
the image, not for the lack of long enough baselines but for lack of
ability to acquire fringes on baselines beyond the first lobe of the
object visibility function.

A way to get around this limitation is known as ``baseline
bootstrapping'' and is discussed in the next section, but use of this
technique imposes limitations on the efficient sampling of the \uv
plane. As a result the requirements for imaging of objects which have
a bright and compact core which remains unresolved when the features
of interest have been resolved are different from those for imaging of
so-called ``resolved-core'' objects like stellar disks. The
requirement for the MROI was that with resolved-core objects it should
be able to make images with 5$\times$5 resolution elements across the
object while with compact-core objects under favourable conditions it
should be able to make 10$\times$10 resel images. This exceeds the
image quality of all existing optical arrays and is comparable to the
images from aperture-masking arrays and many existing radio
interferometers.

Another important aspect of the imaging is the dynamic range of the
images, i.e.~the ratio of the brightest and weakest believable
features in any recovered maps.  This is dependent on a combination of
the Fourier coverage of the observation and the calibration errors on
the visibility amplitudes and phases. Aperture-masking results suggest
that a dynamic range of 100:1 on bright sources is both feasible and
scientifically productive, and so this specification adopted for the
MROI.

\subsection{Sensitivity}
The concept of sensitivity is usually defined for an interferometer in
terms of a limiting magnitude. An object at the limiting magnitude is
just bright enough that it is possible to acquire interferometric data
on the object. This implies that some form of fringe acquisition and
tracking is performed, as without this it is impossible to guarantee
that the interferometer is observing fringes. Thus the limiting
magnitude refers to the overall brightness of the object for
fringe-tracking purposes: if an image with a large dynamic range can
be made, it may be possible to observe objects within the field of
view which are several magnitudes fainter than this.

The extra-galactic component of the top-level science mission for the
MROI sets the basic requirement for its desired sensitivity. At H-band
magnitudes fainter than about 11 the very closest and brightest active
galactic nuclei just start to become visible. However, it is not until
a H-band sensitivity of 14 is reached that of order 100 targets become
visible in the Northern celestial sky. In practice, such a sensitivity
requirement for the array is particularly challenging but later
sections will show that this can be achieved in an optimised design.

\subsection{Wavelength coverage}
The wavelength range from 0.6--2.4\,$\mu$m is key to the science goals
of MROI because it includes the key H$\alpha$ line at 656\,nm and also
the near-infrared range where many of the top-level science targets
are bright. The deterioration in wavefront quality due to atmospheric
``seeing'' means that it becomes rapidly more difficult to compensate
for the atmosphere at shorter wavelengths, while at longer wavelengths
the rising thermal background causes the sensitivity to fall
rapidly. Thus the near-infrared is the regime where the best
interferometric sensitivity can be realised.

\subsection{Spectral resolution}
Spectral resolving powers of $R>5,000$ would be valuable for studying
atomic line emission and absorption in stellar sources but such a
capability would only be useful on relatively bright objects because
of signal-to-noise-ratio issues. A higher priority is a resolving
power $R\simgt 300$ that allows useful isolation of molecular features
from their nearby continuum in stellar sources, and velocity-resolved
imaging in the very nearest AGN, together with a lower-resolution mode
$R\sim 30$ which allows crude spectral diagnostics on fainter targets.

\subsection{Automation and reliability}
It is well established that the most productive optical/infrared 
interferometers have been those with highly automated sequencing and
operation. Our design philosophy for the MROI implicitly assumes
this model, so that tasks such as pre-observing alignment of the optical
trains, self-testing of the detector and beam combination subsystems,
and failure recovery, will all be carried out transparently with a
minimum of operator intervention.

\section{Conceptual design}
\label{sec:conceptual-design}
The requirements for the MROI
emphasise its ability to image {\em faint} and {\em complex} sources in a model-independent manner, and so designing array which optimised both of these aspects was critical. Both of these goals, and especially the sensitivity goal, affect all aspects of the interferometer design and so there is a many-to-many relationship between the elements of the implementation and the design goals. In this section the major elements of the conceptual design are described, the relationship of these elements to the overall goals is described and the trade-offs made between conflicting requirements are addressed. The elements of the design are addressed roughly in order of their importance, but are also ordered by the sequence in which the stellar light impinges on each of these in its journey from the star to being detected as fringes, as shown in Figure~\ref{fig:beampath}.

\subsection{Array configuration}
A critical choice for any interferometer is the number of
telescopes. The cost of an interferometer increases approximately
linearly with the number of telescopes, but the number of \uv points
sampled by $N$ telescopes increases faster than $N^{2}$ . Adequate \uv
coverage is critical to imaging performance, both in terms of image
complexity and dynamic range, but can also be achieved (at the expense
of imaging speed) through repositioning of telescopes.

When imaging resolved-core sources (as defined in
section~\ref{sec:imaging}), an interferometer with a larger number
telescopes is superior not only in imaging speed but also in the
achievable image complexity. This is because the only reliable way to
access baselines which give many resels across such sources is to
utilise the so-called ``baseline bootstrapping'' technique \cite{pauls_bootstrapping_1998}.

In baseline bootstrapping, multiple telescopes are arranged in a
``chain'' such that the nearest-neighbour telescopes are spaced by a
distance which is less than the size of the main lobe of the
visibility function.  On these shortest baselines the fringe
visibility is high enough to allow fringe tracking and hence the
measurement of atmospheric phase differences. These differences can be
integrated along the chain to track the phase perturbations on the
longest baselines where the fringe visibility is too low to allow
direct fringe tracking. The longest baseline which can be accessed is
limited to $N-1$ times the shortest baseline. Hence the maximum number
of resels across such an object rises approximately linearly with the
number of telescopes. This limit cannot be improved upon by moving the
telescopes.
\begin{figure}
  \centerline{\includegraphics[width=\textwidth]{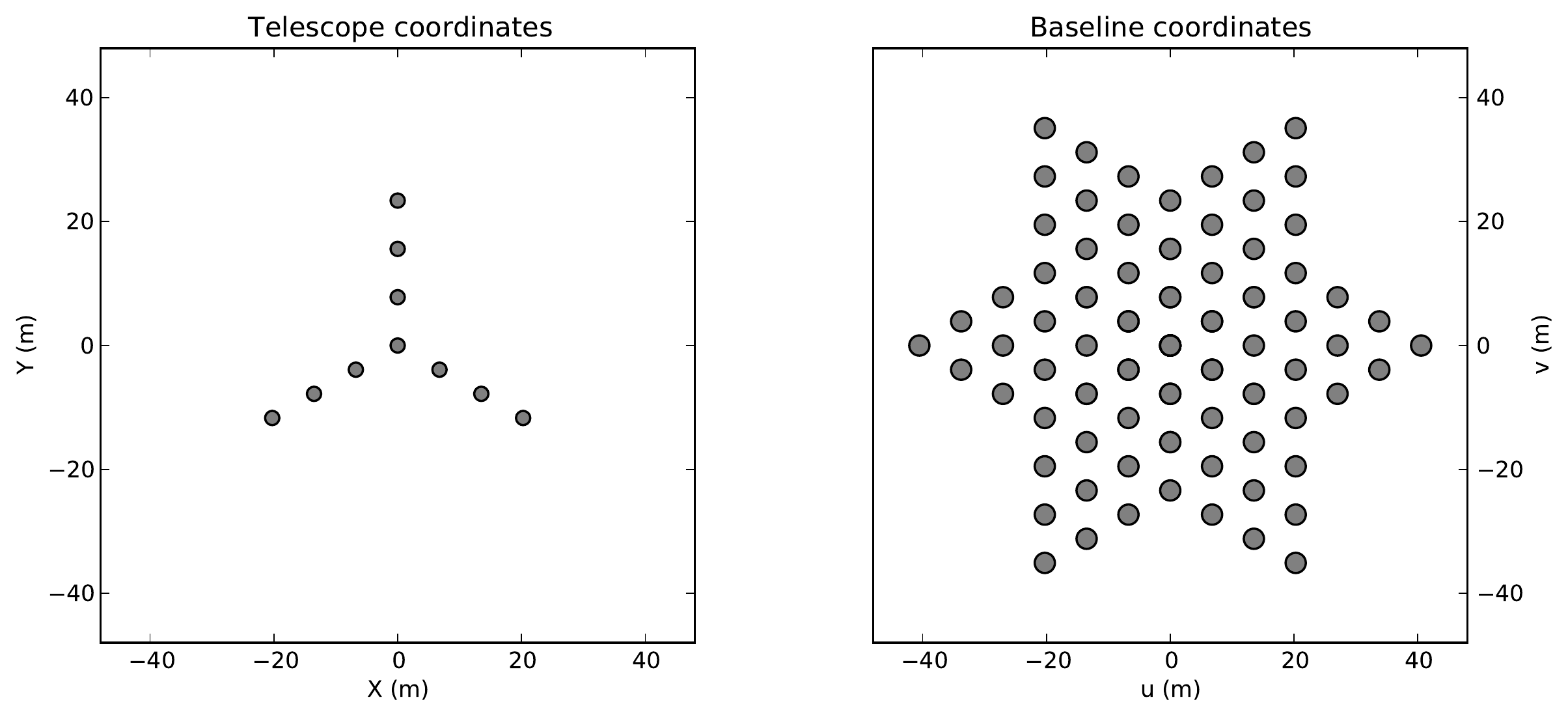}}
  \caption{The ``bootstrapping'' array configuration for the MROI (left) and the corresponding Fourier plane (\uv) coverage (right). The configuration shown is scaled for the most compact baselines but can be scaled up to 340\,m.}
  \label{fig:array}
\end{figure}

The MROI infrastructure is designed for 10 telescopes (more telescopes
could be added but could not be operated simultaneously). When the
telescopes are arranged in an equispaced $Y$-shaped configuration as
shown in Figure~\ref{fig:array} then the array can be viewed as a set
of 3 ``bootstrapping chains'' each consisting of 7 telescopes,
oriented at 120\degree angles to one another. The ratio of the longest
baseline in this array to the nearest-neighbour spacing is 5:1 and so
this will allow approximately 5$\times$5 resel imaging on
resolved-core objects, a capability which is unmatched by any existing
optical interferometer.

At the same time the ``snapshot'' \uv coverage of the array is
superior to any existing optical interferometer, as shown in
Figure~\ref{fig:array}, giving access to 36 well-separated baselines
and 36 closure phases. For compact-core objects, then a non-redundant
configuration can be used, giving access to 45 baselines
simultaneously. Using Earth rotation to increase the density of the
\uv coverage will then allow images with approximately 10$\times$10
resels to be reconstructed on these objects.

A ``Y'' shaped array configuration has been used in a number of
interferometers including COAST, NPOI and CHARA and offers many
advantages including the ability to easily route light from the
telescopes to the centre of the array while using a train of mirrors
whose angles of incidence are symmetric between telescopes. However
the most compelling advantage of this layout for the MROI is that the
array site is on a saddle in the mountain which allows the arms of a
horizontal ``Y'' shaped array to be built with a minimum of earth
movement.

The longest practicable baseline using this layout is approximately
350 meters and this gives a maximum angular resolution of 0.3
milliarcseconds at a wavelength of 600\,nm.

A single configuration of the array gives access to minimum and
maximum angular scales with a ratio of only 5:1, whereas the range of
sizes of possible science targets ranges over more than two orders of
magnitude. The telescopes are therefore designed to be relocatable
between a number of different arrays. The primary array configurations
are scaled versions of the ``bootstrapping'' array, with maximum
baselines ranging from 40\,m to 346\,m and minimum baselines ranging
from 7.8\,m to 67\,m. The array therefore allows access to angular
scales with a range of 44:1.

Four ``bootstrapping'' configurations span this range with a scaling
of approximately two between successive configurations. Different
configurations have been arranged to re-use some of the telescope
stations (by distorting the regular spacing of the telescopes along
the arms by a few percent) so that a total of 28 stations is required.
\subsection{Telescopes and adaptive optics}
One critical factor in the sensitivity of an interferometer is the
telescopes, called ``unit telescopes'' here to distinguish them from
the aperture synthesis telescope comprising the entire array. Large
unit telescopes equipped with high-order adaptive optics (AO) would
appear at first sight to be the most promising route to high
sensitivity, but not only is this an expensive option for an imaging
array with many telescopes, but it also conflicts with the need to
pack the telescopes close together in order to sample larger-scale
angular structure and thereby access targets such as nearby evolved stars and
geosynchronous satellites.

In addition, high-order adaptive optics are only effective if there is
a bright enough reference available to sense the wavefront
perturbations. The angular density of bright natural references
(stars) is sufficiently low that the science target itself is most
often the AO reference. Typical AGN-type targets with an
near-infrared magnitudes of 14 have visible-wavelength magnitudes of
around 16 and these are too faint to drive current AO systems at
radial orders higher than tip-tilt \cite{wilson_adaptive_1996}.

Tip/tilt correction gives the best limiting magnitude on telescopes
which are of order 2--3$r_{0}$ in diameter \cite{buscher1988} and so,
in the absence of laser guide star systems at each telescope,
telescopes in the 1--2 metre size range provide close to the best
possible interferometric sensitivity at optical and near-infrared
wavelengths.
\subsection{Beam relay}
\begin{figure}
  \centerline{\includegraphics[width=0.8\textwidth]{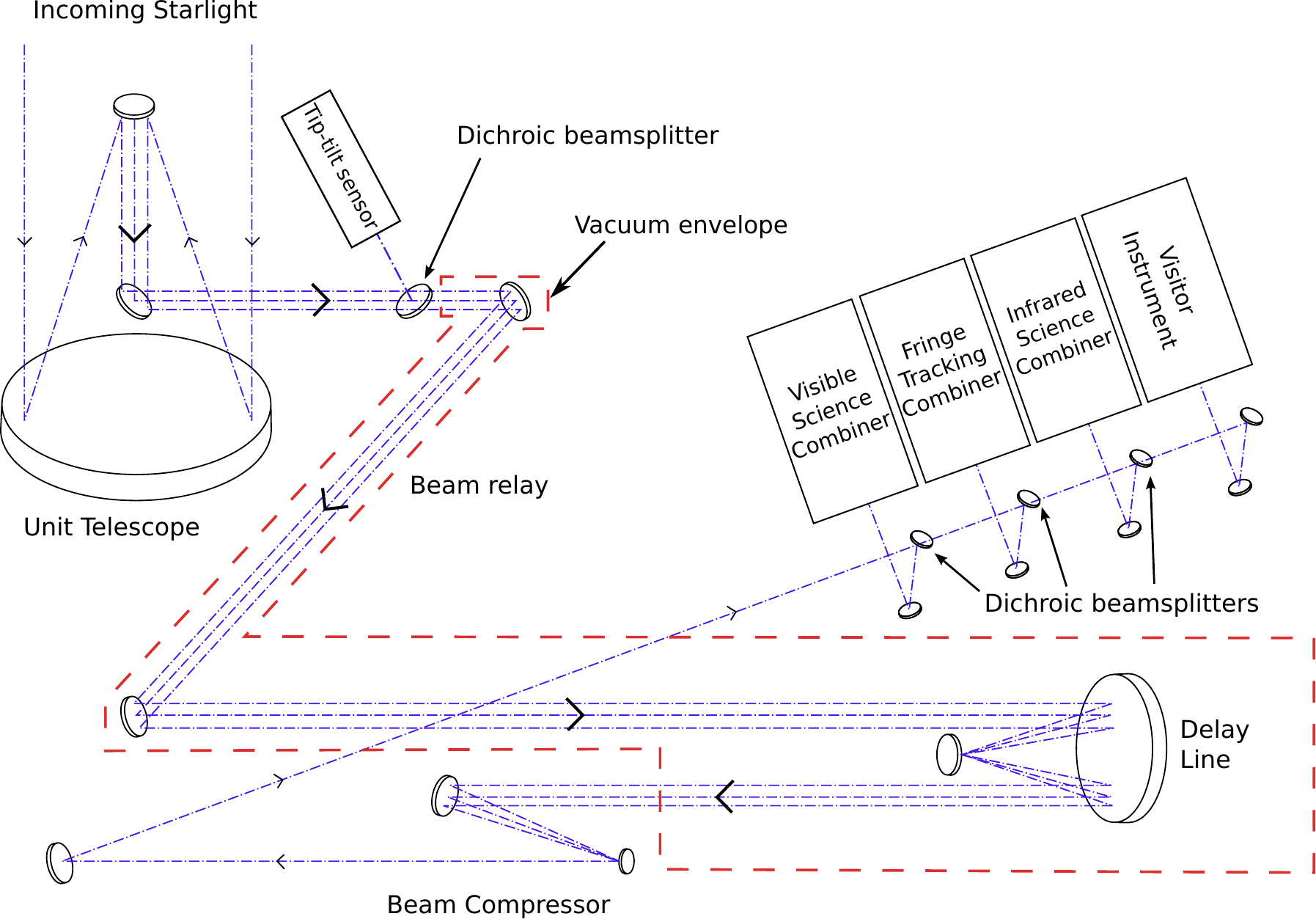}}
  \caption{The beam path from one of the telescopes in the array to the beam combiner}
  \label{fig:beampath}
\end{figure}

The parallel beams exiting from the unit telescopes are sent to the
beam combining optics as a set of parallel beams, where most of the
path is in vacuum, as shown in Figure~\ref{fig:beampath}. Vacuum beam
propagation was chosen over monomode fiber optics for beam transport
as it allows the propagation of the entire bandpass from the optical
to the infrared to be handled in a single system rather than requiring
splitting the bandpass into multiple bands for propagation, which
inevitably incurs additional losses.

The diameters of the beams need to be significantly larger than the
Fresnel zone size $\sqrt{\lambda z}$ in order to minimise the effects
of diffraction for propagation over a distance $z$. For propagation
distances of order 1\,km and a wavelength of 2.2\,$\mu$m, this zone
size is of order 5\,cm and so relatively large optics are required for
propagation over these distances. The propagation distances inside the
beam combining area are of the order 20\,m so smaller beams can be
used in order to keep the size and cost of the more complex beam
combination optics within reasonable bounds. A set of beam compressors
between the delay lines and the beam combination optics serves to
transform between these two beam sizes.
\subsection{Delay lines}
Dynamical optical path compensation is required in order to equalise
the path-lengths travelled by the starlight from the target to the
detector. The path-length stroke required from these compensators
scales with the size of the array, and arrays of comparable size such
as CHARA and NPOI utilise a two-stage path compensation system,
consisting of a switchable delay to give large stroke and a
continuously-variable delay to give fine control of the
path-length. The approach adopted at MROI was to have a single-stage
system which introduces all the delay in a continuously-variable
manner. This minimises the number of reflections needed and at the
same time avoids the switching overheads associated with the two-stage
systems.
\subsection{Science instruments}
The 10 beams exiting from the delay lines are compressed and then
spectrally split using optimised dichroics
\cite{hobson_markov-chain_2004} between a number of interferometric
beam combiners. A visible-light combiner and a near-infrared (JHK)
combiner can be operated simultaneously, and space has been left for a
``guest'' instrument which might substitute for one or other of these
combiners. These beam combiners will be optimised for operation on
faint sources, but at low light levels the signal-to-noise ratio of
the fringes decreases as the number of beams which are combined
simultaneously is increased, because the photons from all the
telescopes contribute to the noise on all the baselines. Therefore the
MROI science instrument concepts are based on a number of parallel
combiners, each of which combine a different subset of the beams. A
beam ``switchyard'' then allows the beams to be ``shuffled'' between
combiners to allow all pairs of telescopes to be interfered with one
another. One possibility would be to have two combiners accepting 4
beams each. This would produce fringes on 12 baselines and 6
independent closure phases. A series of 4 reconfigurations of the
switchyard would allow access to all 45 baselines and 24 of the 36
independent closure phases possible from the array.

\subsection{Fringe tracking}
Fringe tracking is required to overcome effects of the random
path-length perturbations introduced by the atmosphere and the
instrument. Fringe ``cophasing'' attempts to compensate for the motion
of the fringes at the sub-wavelength level while ``coherencing'' is
coarser and attempts to reduce the fringe motion to less than the
coherence length of the light, which can be many microns.

Fringe tracking in the MROI will mostly utilise the group delay method
which is based on observing the phase differences between the fringes
in multiple spectral bands. This allows fringe coherencing on sources
about 10 times fainter than is possible with cophasing methods
\cite{dfbphd} and so is the best way of achieving the MROI
faint-science goals. The MROI has been designed with a separate
fringe-tracking combiner rather than using the science instrument to
derive a fringe-tracking signal, because it allows each combiner to
be optimised for a different role.

\subsection{Alignment}
Misalignments of components of the interferometric beam train are a
source of wavefront error in an interferometer which can lead to
significant losses in light throughput and fringe visibility. Many of
the components need to be realigned on at least a nightly basis to
account for drifts which have occurred during the day. A significant
fraction of the realignment in many interferometers requires manual
intervention and this can limit the amount of realignment which can be
done at the start of the night.  The MROI was designed from the start
to allow automated alignment of the majority of the optical train in
order to minimise the time overhead and to increase the overall
accuracy of the alignment.

\subsection{Control software}
Automated operation of an interferometer is critical to its scientific
productivity, as large numbers of subsystems must work together to
produce the interference fringes, and the cadence between observations
is necessarily short in order to allow for observations of calibrator
stars as close in time to the target observations as possible. A
second critical feature for interferometers is continuous recording
and storing of engineering data, as the calibration of the fringe
visibility can depend on many variables within the system, and some of
these can only be determined after the fact. For commissioning and
operations, near-real-time display of data from multiple subsystems on
the same console is essential to debugging operation of the system.

The control software concept is based around independently-developed
sub-systems forming a distributed system communicating over
Ethernet. The system as a whole is only soft-real-time as the
hard-real-time elements are confined mostly within subsystems such as
the fast tip/tilt system. Real-time communication between subsystems
is needed only between the fringe-tracker and the delay lines and this
can be provided by a dedicated communications link.

\section{Error budget}
\label{sec:error-budget}
Given that increasing telescope size does not provide substantial
gains in sensitivity, the overall efficiency of the system becomes
paramount.  In order to achieve the faint-object science goals for the
MROI a two-fold strategy to achieve the maximum optical efficiency was
adopted. The first component of the strategy was to focus on an
optical design which was as simple as possible, so as to minimise the
number of optical elements in the system. In some cases this meant
sacrificing capabilities, for example a full polarimetric capability,
which could have offered increased science performance on brighter
targets. It should be noted that the polarisation fidelity of the
interferometer (defined as the ability to faithfully measure the
object morphology in the Stokes I component) was not sacrificed as
this did not involve compromising the optical efficiency of the system
\cite{buscher2009minimizing}.

This resulted in a beam train which has a low number of optical
elements compared with many existing arrays: in the MROI design
starlight experiences 13 reflections from the entrance of the
telescope to the entrance of any of the beam combining
instruments. This can be compared with the VLTI, where the light
experiences 32 reflections between the equivalent locations in the
beam train \cite{vlti-icd}.

The second component of the strategy was to make sure that each
surface introduced the minimum optical loss consistent with an
affordable budget and minimal technical risk. This can be achieved in
principle by only using components with the best possible coatings and
which are manufactured to the tightest possible wavefront
tolerances. However, the components tend to cost exponentially more
the tighter the tolerances are set, and achieving these tolerances is
easier for some components than for others. From this emerges the
concept of an optical ``error budget'', in which global values for the
allowable losses are defined and these budgets are then shared between
subsystems in a way that excessive requirements are not placed on any
single subsystem.

In interferometric instruments, both the losses to the number of
photons and losses that reduce the fringe visibility are
important, and so both must be considered as part of the optical error
budget. The majority of the visibility loss is caused by wavefront
errors, including both temporal ``piston jitter'' errors caused by
vibrations in the system and spatial wavefront errors such as tilt and
focus errors. As a simplifying assumption, the fringe visibility loss
$\gamma$ is assumed to vary as
\begin{equation}
\label{eq:1}
\gamma=\exp(-\sigma_{\rm diff}^{2}/2)
\end{equation}
where $\sigma_{\rm diff}$ is the spatial or temporal RMS difference of
phase between the interfering wavefronts. This assumption is related
to the ``extended Mar\'echal approximation'' used in aberrated optical
systems and empirically found to be a reasonable approximation for
$\sigma\simlt 2$ \cite{mahajan_strehl_1983,hardy_adaptive_1998}. If
each beam train in the interferometer introduces uncorrelated phase
errors with RMS value $\sigma$ then
\begin{equation}
\label{eq:2}
\gamma=\exp(-\sigma^{2}).
\end{equation}
If in addition each component labelled $i$ in the beam train
introduces a spatial or temporal wavefront error $\sigma_{i}$ which is
uncorrelated between components then the total error is the root
summed squared (RSS) of the individual errors
\begin{equation}
\label{eq:3}
\sigma^2=\Sigma_{i} \sigma_i^2.
\end{equation}
As a result, the visibility loss contributed by different components
can be combined multiplicatively in the same way that losses in photon
throughput can be combined.

\begin{figure}
  \centerline{\includegraphics[width=0.8\textwidth]{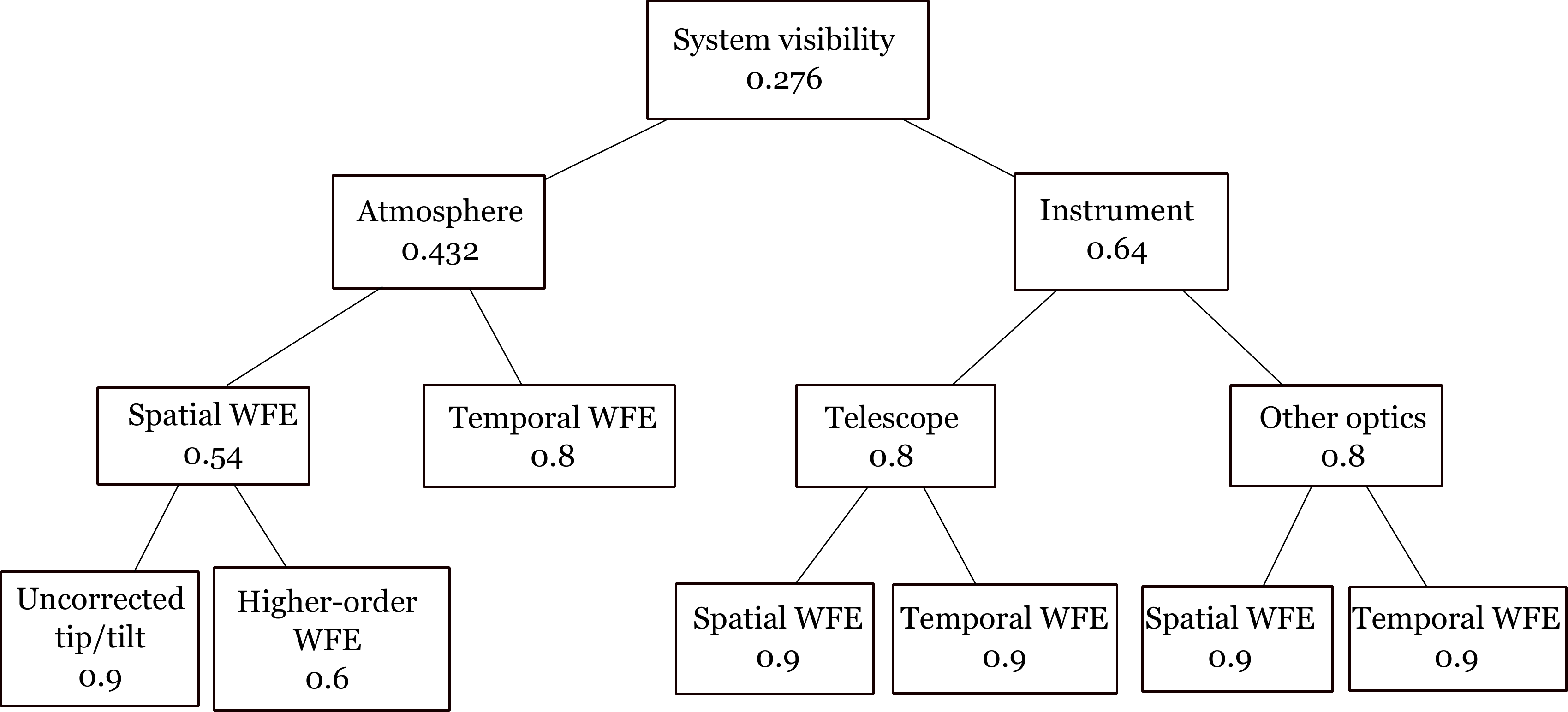}}
  \caption{Part of the initial error budget for the MROI, showing the
  top-level allocations of the wavefront error to different
  subsystems. For simplicity, only a subset of the branches of the
  error budget ``tree'' are shown. All quantities are for a wavelength
  of 1600\,nm, \ie the centre of the astronomical H-band. The abbreviation WFE stands for ``wavefront errors''.}
  \label{fig:errorbudget}
\end{figure}

Figure~\ref{fig:errorbudget} shows an outline of the visibility loss
budget for the MRO interferometer. This is shown in a ``tree'' form
showing how the budget is split amongst various components. The loss
is shown for a wavelength of 1.6\,$\mu$m which is the wavelength of
the highest-sensitivity fringe-tracking for MROI. The visibility
losses in the error budget can be broken down into factors due to
atmospheric seeing and those due to the instrument. At the reference
wavelength, the MROI telescopes have a diameter $D/r_{0}\approx 2.4$
in the reference seeing conditions ($r_{0}=14$cm at a wavelength of
500\,nm) and so with perfect tip/tilt correction the RMS visibility
loss factor due to the uncorrected higher-order atmospheric
aberrations is about 0.6 \cite{buscher1988}. Residual tilts caused by
the imperfect performance of the tip-tilt correction system are
budgeted to allow an additional visibility loss factor of
0.9. Assuming an exposure time of 2$t_0$ then the visibility loss due
to atmospheric temporal fluctuations is about 0.8 \cite{buscher1988}.

Spatial and temporal wavefront errors introduced by the interferometer
optics are budgeted to each introduce a reduction in the fringe
visibility by a factor 0.8, values comparable to the atmospheric
losses and corresponding to $\lambda/14$ RMS wavefront errors in each
case. The telescope optics are the largest elements in the optical
train and the most prone to vibration, so half of the RSS wavefront
error budget (corresponding to a 10\% visibility loss) is allocated to
the telescopes and the remaining half to the rest of the
interferometer subsystems.

The remaining optical train has smaller optics, but a larger number of
components. Each optical element will introduce a wavefront error due
to imperfections in its manufacture, but in addition the curved
optical components can introduce wavefront errors due to alignment
errors such as defocus. Assigning the wavefront errors equally to
manufacturing errors in each element and the alignment error leads to
the spatial wavefront error budget being shared between more than 20
contributors. As a result, each contributes on average only 0.5\% or
so to the final visibility loss.

This level of loss corresponds to an RMS wavefront error of
$\lambda/90$ at the fringe-tracker wavelength of 1600\,nm,
corresponding to $\lambda/35$ RMS at the 633\,nm HeNe laser wavelength
at which optics are normally tested. Small optics can be routinely
polished to give a wavefront quality of $\lambda/10$ peak-to-valley
and the RMS wavefront error will typically be a factor of 5 or so
smaller than the peak-to-valley \citep{Porro1999}, so a $\lambda_{\rm
HeNe}/10$ mirror will contribute a visibility loss well within the
typical tolerances assigned.  Thus it can be seen that by using an
error budget, the wavefront error requirements can be shared out such
that no single component has unfeasible requirements placed on it.

An error budget for both photon loss and visibility loss was developed
for the entire system with a starting goal of an overall visibility
loss factor of $\gamma=0.276$ and a photon throughput from the top of
the atmosphere to detected photoelectrons of 20\%. Adjustments to the
error budget based on more detailed designs of the subsystems systems
has resulted in a predicted system visibility loss factor of
$\gamma=0.30$ and a photon throughput of 13\%.

\section{Limiting magnitude}
\label{sec:limiting-magnitude}
The science requirements lead to a specification that the interferometer should allow fringe tracking on an object with an H magnitude of 14 which is unresolved on the nearest-neighbour ``bootstrapping'' baselines. With a 13\% throughput and assuming a pairwise, nearest-neighbour
fringe-tracking beam combiner, 1540 photons/second will be detected over the H-band from such an object in
each fringe pattern, and
so in a 35\,ms exposure (corresponding to 2$t_{0}$ in the reference
seeing conditions of $t_{0}=4.4$\,ms at a wavelength of 500\,nm),
approximately $N=54$ photons from the target will be detected. With an RMS system visibility of $\gamma=0.30$, the fringe signal level (sometimes called the
coherent flux) will therefore be
\begin{equation}
\label{eq:4}
S=\tfrac{1}{2}\gamma N=8.1\,{\rm photons}.
\end{equation}

There will be an additional 8.2 photons detected per frame of sky and
thermal background photons giving a photon noise level of
\begin{equation}
\label{eq:6}
\sigma_{\rm phot}=\sqrt{62.2}=7.9\,{\rm photons}
\end{equation}
Assuming that the fringe tracker uses 4-bin ``ABCD'' sampling and 5
spectral channels across the H-band, then the noise due to detector
read noise is
\begin{equation}
\label{eq:5}
\sigma_{\rm read}=\sigma_{\rm pix}\sqrt{20}=8.4\,{\rm photons}
\end{equation}
where $\sigma_{\rm pix}$ is the noise on the readout of a single
pixel, assumed to be 2 electrons: near-infrared detectors with
single-read noise values of this order are now available.  Thus the
signal-to-noise ratio for a fringe measurement in a single exposure
will be
\begin{equation}
\label{eq:7}
{\rm SNR}=\frac{S}{\sqrt{\sigma_{\rm phot}^2+\sigma_{\rm read}^{2}}}=0.70
\end{equation}
Simulations of fringe tracking with this number of spectral channels
\cite{dfbphd} show that group-delay tracking is possible when the
signal-to-noise ratio per exposure is as low as 0.56, so the MROI
science goal of fringe-tracking on a H=14 AGN appears to be
achievable with this design. Improved margins could be obtained if
detector read noise performance better than 2 electrons can be
achieved using multiple non-destructive reads, as has already been
demonstrated in the laboratory \citep{finger_evaluation_2012}.

\section{Experience from implementing the design}
\label{sec:lessons-learned}
Because the MROI is not currently operational it is not yet possible
to provide definitive evidence that by implementing the conceptual
design all the scientific performance goals will definitely be met.
Instead we can look at what the experience of designing (and in many
cases building and testing) the subsystems on the basis of the
conceptual design tells us about how well the conceptual design stands
up to the reality of implementation. Here we discuss this experience
for selected subsystems. The aim is not to describe the designs of
these subsystems in detail, as this has been (and will be) discussed
elsewhere, but rather to comment on where the design was unusual and
where there were technical roadblocks due to unrealistic constraints
imposed by the conceptual design.
\subsection{Unit Telescopes}
\begin{figure}
  \centerline{\includegraphics[width=0.7\textwidth]{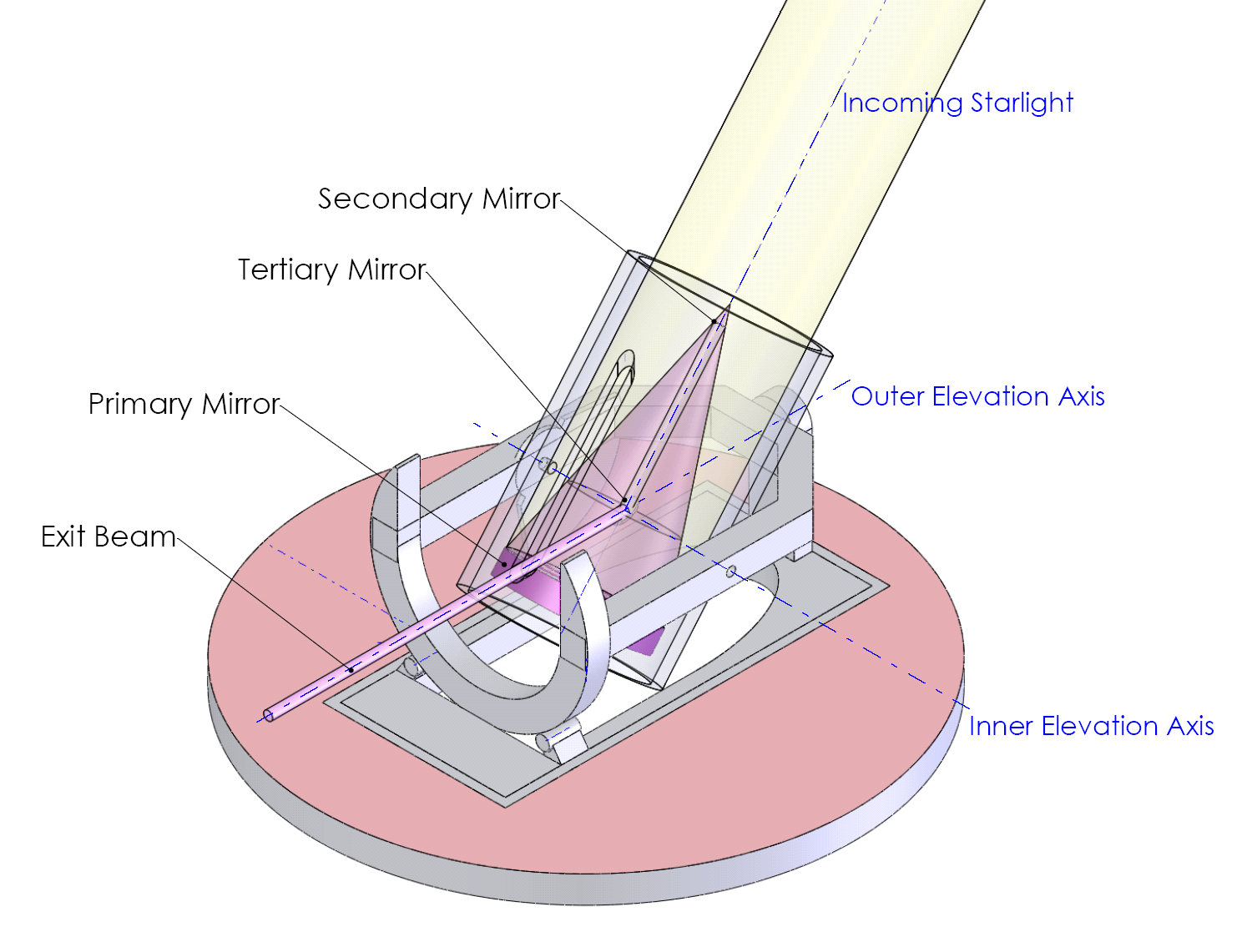}}
  \caption{Schematic of the three-mirror ``elevation over elevation'' (or ``alt-alt'') telescope design used for MROI.}
  \label{fig:elel}
\end{figure}
The unit telescopes for the MROI are 1.4\,m in diameter and use a
relatively unusual mount design in order to improve light throughput to
the interferometer.  This ``alt-alt'' design as shown in
Figure~\ref{fig:elel} allows the starlight to be captured and
converted into a narrow collimated beam travelling in a fixed
horizontal direction using only 3 reflections. This therefore has
lower reflection losses than conventional ``alt-az'' mount which
typically requires 7 reflections to achieve the same result.

In addition, the telescope has been specified with a high wavefront
quality in order to reduce fringe visibility losses due to the
optics. Wavefront quality for an interferometric telescope includes
also the ``piston term'' errors which can be induced by telescope
vibrations and also pupil wander which can cause the beams from
different telescopes not to overlap.

The telescope mounts have been built and tested in the factory by AMOS
(Advanced Mechanical and Optical Systems) and have passed all major
performance tests. The OPD vibrations are less than 40\,nm RMS over a
35\,ms exposure and the pupil motion is less than 350\,$\mu$m in
radius.

The telescopes are designed to be relocatable between different
stations along the array. The concept adopted is to relocate the
telescope and its enclosure together, using a transporter based on
those used for stacking shipping containers.

Despite the relatively small size of the telescopes, attaining the
shortest telescope spacings in a close-packed configuration proved
difficult. The enclosures were designed to allow only the minimum
space for support electronics, servicing and relocation, and the
minimum spacing achieved was 7.8\,m, only just short enough to overlap
with the baselines offered by 8-meter-class telescopes.
\subsection{Fast Tip-Tilt Systems}

The location of the fast tip-tilt (FTT) systems in the MROI beam-train
is perhaps unusual; we have opted to place the sensors on the Unit
Telescope Nasmyth platforms (with corrections applied by the UT
secondary mirrors). Compared with locating the sensors in the beam
combining building, this increases the photon flux available for
sensing atmospheric tip-tilt perturbations and thus enhances the
limiting sensitivity of the fast tip-tilt systems. This
sensitivity limit would otherwise be the limiting factor for
observations of red objects such as AGN and YSOs. With the sensor at
the telescope, we expect to achieve a residual two-axis tip-tilt error
of 60.8\,mas RMS on the sky at $m_{V}=16$ in the reference $0.7^{\prime\prime}$
seeing, by taking advantage of the high quantum efficiency and
sub-electron read-noise offered by electron-multiplying CCD detectors.

However the disadvantage to this location is the sensitivity to non-common-path errors, whether movements of the FTT optics which shift the image
on the tip-tilt sensor or changes in the alignment of the beam relay mirrors which are not be
seen by the tip-tilt sensor. The system has therefore been designed for high
optomechanical stability
($0.015^{\prime\prime}$ on the sky for a $5\,^{\circ}$C change in
ambient temperature, corresponding to a 0.5\,$\mu$m image shift on the
tip-tilt sensing CCD). The design of
the FTT systems, including optomechanical considerations, is discussed
in more detail by \citet{young_mroi_2012}.

The level of intra-night stability we expect to achieve with the final
design, based on tests of prototype components, is a factor 2--4
greater than originally budgeted. To accommodate this, as well as any
instability in the tilts of the beam relay mirrors, we have designed a
continuous alignment system to be installed in the beam combining
building --- this is described in section~\ref{sec:alignimplementation}.

\subsection{Beam Relay System}
The diameters of the beams for propagation from the telescopes to the
BCA was chosen to be 95\,mm based on beam propagation studies which
included the effects of atmospheric turbulence
\cite{horton_diffraction_2001}. Inside the BCA the beam is compressed
to approximately 13\,mm.

Due to the Y-shaped array design, only two reflections are needed
between the telescopes and the delay lines, but the mirrors involved
need to be outdoors with their centres 1.6\,m above ground level. The
concrete piers to hold these mirrors needed to be engineered to
achieve the appropriate stability in the face of wind loads, and this
proved technically challenging, resulting in bulky and costly
piers. Although the mirror mounts will be inside vacuum enclosures,
they and their piers will be subject to substantial diurnal
temperature fluctuations and so the stability of the beam relay is
being tested to see if they will be sufficiently stable during the
night.

\subsection{Delay Lines}
\begin{figure}
  \centerline{\includegraphics[width=\textwidth]{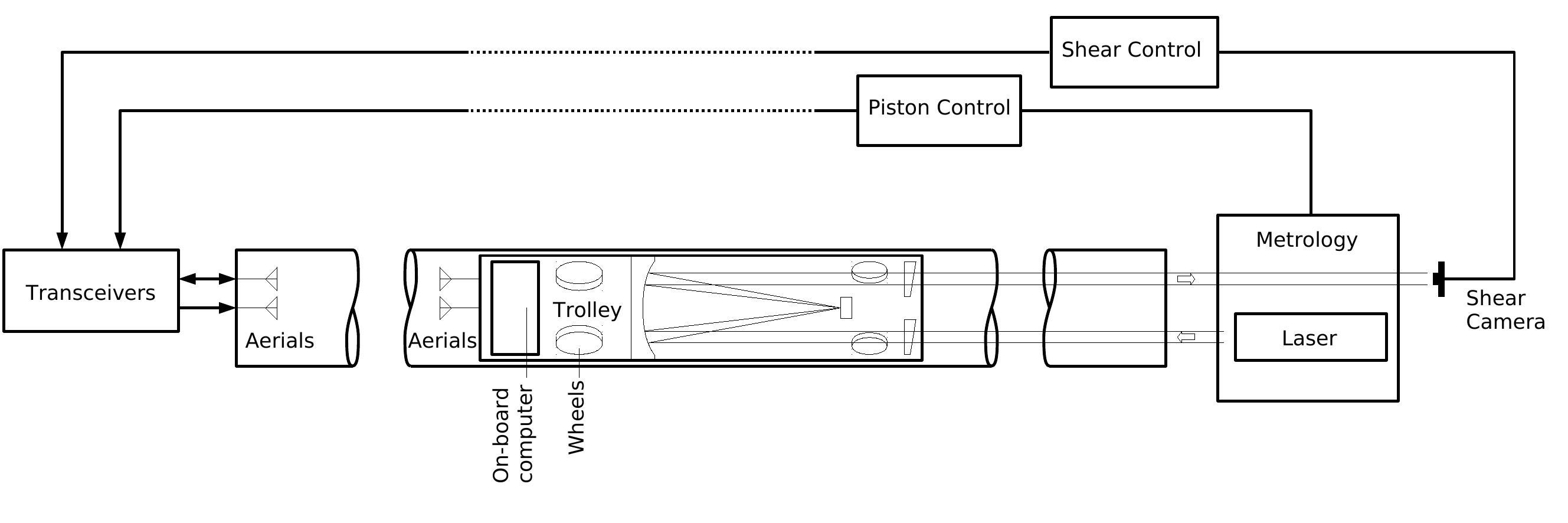}}
  \caption{Conceptual block diagram of the delay line. The path for the metrology beams is shown in this plan view --- the science beams are not shown.}
  \label{fig:delayline}
\end{figure}

The desire to deliver high throughput and efficient
operations were major drivers for the implementation of the
MROI delay lines. The detailed design of these has been presented
elsewhere \citep{Fisher2010} and so here we
summarise only a few of their novel features.

The MROI design supports a delay range from 0
to 380\,m, in vacuum, and is realised in a single stroke with only
three reflections. As a result there is no need for any separate
switchable ``long'' delays nor any longitudinal dispersion correctors.

Implementing this functionality required a design which concentrated in making the delay scalable to hundreds of metres at an affordable cost. This involved innovations such as using the vacuum vessel as the surface on which the delay carriage runs, wireless control and inductive supply of power so that there are no cables to be dragged as the optical
delay is adjusted. Figure~\ref{fig:delayline} shows a schematic of the delay line, indicating that in addition to the normal control loop for the position of the catseye (controlling the ``piston'' component of the wavefront), the delay lines incorporate active control of the shear of the return beams, to allow the use of less straight and hence more affordable pipes for the vacuum vessel. 

Tests in Cambridge in a 25\,m long evacuated test rig have
demonstrated that the system can meet all the system's top level performance requirements,
including a delay jitter $< 15$\,nm RMS over 10\,ms; $< 41$\,nm RMS
over 35\,ms, and $< 55$\,nm RMS over 50 ms ($\lambda/40$ RMS over
$2t_0$). Full repositioning of the
carriage over the whole OPD range can be executed in 5\,min. The first delay
line carriage has been delivered to the MROI site and site acceptance
tests in a 100\,m long delay line will take place later this year.

\subsection{Fringe Tracking Beam Combiner}

The fringe tracking beam combiner that has been adopted for the MROI
(Infrared COherencing Nearest Neighbour tracker: ICONN) is a so-called
nearest-neighbour design in which the beam combiner optics only mix
beams from pairs of telescopes that are closest to each other. This
arrangement maximises the per-baseline signal-to-noise ratio. 

The design of ICONN \citep{Jurgenson2008} has allowed for up to 10 input beams, from the
three arms of the MROI, and straightforwardly manages the unusual
situation of the telescope at the vertex of the array which, unlike
all the other array elements, has three nearest-neighbours. Currently
the performance of ICONN is being tested in a single-baseline
demonstration in the laboratory using a PICNIC detector to sense the
dispersed fringes. Initial results are promising, and show very high
fringe visibilities and an excellent level of stability in a
relatively uncontrolled environment.

\subsection{Science Beam Combiner}

The eventual implementation will
include two science beam combiners, using ``visible''
(0.6--1.0\,$\mu$m) and near-infrared ($J$, $H$, $K$) light
respectively. A decision was made to prioritise development of the
near-IR science beam combiner over the visible-wavelength one, as this
enables a larger fraction of the science mission and places less
stringent demands on the performance of the rest of the
interferometer.

A range of design concepts for the near-IR science beam combiner were
considered. These were described in \citet{Baron2006}.  
All the designs include a fast optical ``switchyard'' would comprise plane mirrors mounted
on precision slides to enable the rapid selection of beams, such that
different subsets from of the available beams are selected for
entry into the combiner.  This approach maximizes the instantaneous
fringe signal-to-noise but incurs a small efficiency overhead
associated with reconfiguring the switchyard every few minutes to
measure all of the baselines. Initial tests of candidate slides
suggest that it will be possible to use a look-up table to correct for
errors in the mirror orientation following each reconfiguration.

\subsection{Alignment}
\label{sec:alignimplementation}
The biggest sources of alignment error in the interferometer are due to drifts in the tilts
of mirrors in the beam train, which leads to tilt errors in the
propagating beam. Over long distances these tilt errors lead to errors
in the transverse position of the beam (``beam shear'') and so control
of these errors requires measuring and correcting both the tilt and
the shear simultaneously.

\begin{figure}
  \centerline{\includegraphics[width=0.6\textwidth]{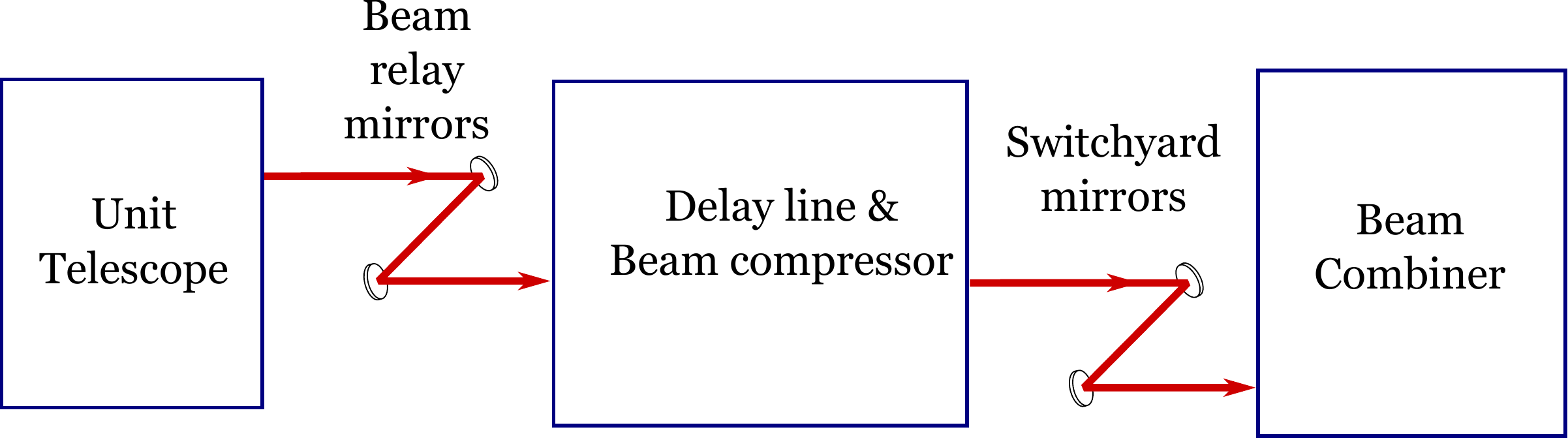}}
  \caption{Block diagram of the alignment configuration for one beam path, showing the mirror pairs used for adjustment of tilt and shear.}
  \label{fig:alignment}
\end{figure}

For the purposes of nightly alignment, the optical train from the
telescope to the beam combiner can be conceptually split into three
independent optical ``sub-trains'' whose optical axes must be aligned
with one another, as shown in Figure~\ref{fig:alignment}. The optical
axes of these sub-trains are defined by the inner rotation axis of the
unit telescopes, by the line of motion of the delay line and by the
optical axes of the relevant beam combiner respectively. These axes
are connected by pairs of mirrors which can be adjusted to align the
tilt and shear of the axes between sub-trains.

The alignment of these sub-trains will be accomplished by
sending out a reference ``pilot beam'' which is aligned with the delay
line axis and which shines both out towards the telescopes and back
towards the beam combiners.
Tilt and shear sensors at the telescopes and the beam combiners will
allow the pilot beam position and direction to be compared with
the corresponding optical axes. Automated procedures will be used
to adjust the appropriate relay mirror pairs for any
misalignment \citep[see][]{Shtromberg2010}. Most of this activity will be automated, and so is
expected to run in parallel on the beam trains for different
telescopes, minimising the overhead for system alignment.
 
For this strategy to be effective, most of the alignment system
components will be located in thermally-stable environments so that
drifts during and after the alignment process are kept to a minimum.
At the MROI the beam combining area (BCA) is stabilised so that
diurnal temperature fluctuations are at the 0.1\degree C level and
this zone will contain the optics for the pilot beam injection as well
as all the beam compressors and the beam combination optics. The delay
lines are less susceptible to thermal drifts and so are housed in a
passively stabilised enclosure, the delay line area or DLA. Over the
seasons the DLA will be allowed to drift in temperature by more than
10\degree C but on a diurnal basis the temperature change is of the
order 1\degree C, and so intra-night drifts will be kept within an
acceptable range.

An issue which was recognised at a late stage in the design was that during the night
(\ie after the automated alignment) it will be likely that
temperature-induced perturbations of the components of the beam trains
mounted either on the telescope's Nasmyth tables or the ``exposed''
beam relay mirrors will lead to slow ($>100$\,s) drifts in
alignment. To address this, there will be a secondary low-bandwidth
alignment system located within the temperature-controlled BCA. This will
correct slow changes in alignment by picking off light in the
940--1000\,nm bandpass and monitoring the tilt and shear drifts in real-time.

\subsection{Controls}
In implementing the MROI control system, we have had to solve one
overriding issue -- the need to integrate many diverse systems into an
integrated whole that can be operated efficiently. The diversity
arises for a variety of reasons, including intrinsic differences in
the kinds of functionality provided by the sub-systems (such as
capturing sensor readings, closing fast servo loops, and complex
algorithms for e.g.\ alignment or fringe tracking), and implementation
constraints (such as the need to interface specific hardware devices
or run under certain operating systems).

The problem of merging these systems is solved by using standardised
interface software that is automatically generated from a simple
high-level description of these systems. The generated interface code provides functionality for object
construction and destruction, system configuration using data obtained from the
central database, receiving commands, publishing monitor data, faults
and alerts, and subscribing to monitor data published by other
interferometer systems.

The control system is capable of managing independent subsets of the
interferometer array. A typical use of this feature would be to
operate most of the unit telescopes as a single system, while
simultaneously doing calibrations on two recently moved
telescopes. For additional information see \citet{Farris2010}.

\section{Conclusions}
\label{sec:conclusions}
Although the MROI is not yet operational, a number of conclusions can be drawn from the design of the interferometer. The first of these is the importance of imaging to the scientific productivity of interferometers and the technical implications that arise from this. It is clear from the science case for the MROI that there are a large number of science targets for which imaging would provide critical new insights. All these targets are complex, and so the number of degrees of freedom in any model used to describe the target is large. Model-independent imaging is the only way to constrain these degrees of freedom in a way which can constrain which models are appropriate and is robust to any degeneracies in the models.

Given the scientific importance of imaging, we have argued that having a large number of telescopes is critical in meeting the imaging goals, not only because of the substantial increase in the speed of imaging, but because for many objects the problem of tracking atmospheric phase perturbations can only be tackled with a ``bootstrapping'' array, which necessarily requires many telescopes. The MROI with 10 telescopes will be able to image ``resolved-core'' objects with of order $5\times 5$ resels across the image; to make more detailed images of these objects will require correspondingly more telescopes --- moving the telescopes around will not be able to achieve this effect.

A second scientific focus of the MROI is that of sensitivity. Our analysis suggests that increasing the aperture size of the unit telescopes in the interferometer will have limited effect in increasing the faintness of the targets which can be observed, and that concentrating on designing an efficient beam train which is able to provide the required functionality may have greater benefits. Our analysis has assumed the use of natural-guide-star adaptive optics, and the availability of cheap and reliable laser-guide-star technology could change this picture. Even in this latter scenario the increase in sensitivity can only be realised with larger and so more costly telescopes and this needs to be tensioned against the smaller number of telescopes which can be deployed within a given budget. 

A final remark can be made about the process of the development of the conceptual design which is reflected in the structure of this paper. It is often argued that to design a new interferometer requires first the development a ``killer'' science case, and this then drives the technical development of the interferometer. The process described here is a more nuanced one which starts from an analysis of the scientific successes and shortcomings of existing arrays, followed by a judgement of which of the shortcomings can be overcome most fruitfully without requiring inordinate amounts of technical development. Having established the likely direction of technical evolution, the next step is to establish if there is indeed a science case that would capitalise on this increased technical capability and to develop this case more fully. Finally, this science case is used both to justify the requests for funding and to guide the design of the interferometer. Thus, and we believe this observation is true of many fields and not just interferometry, a successful instrument development process is in reality not a linear and purely ``science-driven'' one, nor is it a ``technology-driven'' one, but instead the flow between scientific and technical drivers is a never-ending back-and-forth between the two.
\section*{Acknowledgements}

The authors would like to acknowledge the help of E.B.~Seneta and J.~Kern with the diagrams; development of the initial concept was aided by discussions with C. Briand and T. Sauza.

\newif\ifbib
\bibfalse
\ifbib
\bibliographystyle{ws-jai}
\bibliography{jai-mroi-extra,all,dfbpubs,zauto}
\else

\fi
\end{document}